\documentclass[sigconf,authorversion]{acmart}

\usepackage[utf8]{inputenc}
\usepackage[T1]{fontenc}

\usepackage{booktabs} 
\usepackage{longtable}
\usepackage{array}
\usepackage{multirow}
\usepackage{wrapfig}
\usepackage{float}
\usepackage{colortbl}
\usepackage{pdflscape}
\usepackage{tabu}
\usepackage{threeparttable}
\usepackage[normalem]{ulem}

\usepackage{caption} 
\usepackage{graphicx}
\graphicspath{{results/}{results/figures/}}
\DeclareGraphicsExtensions{.pdf,.png,.jpg}
\usepackage{colortbl}

\usepackage[acronym,shortcuts]{glossaries}
\PassOptionsToPackage{hyphens}{url}
\RequirePackage{cleveref}
\usepackage{microtype}
\usepackage{csquotes}
\usepackage{amsmath}
\usepackage{amssymb}
\usepackage{wasysym}
\usepackage{tabularx}
\usepackage{threeparttable}
\usepackage[inline]{enumitem}
\usepackage{listings}
\usepackage{siunitx}
\usepackage{paralist}
\usepackage{flushend}
\definecolor{darkblue}{rgb}{0,0,0.75}
\definecolor{darkgreen}{rgb}{0,0.5,0}

\newcommand{\java}[1]{\lstinline~#1~}
\newcommand{\outcome}[3]{$\overline{#1}_{\text{#2}} = #3$}
\newcommand{\effectiveness}[2]{\outcome{E}{#1}{#2}}
\newcommand{\efficiency}[2]{\outcome{P}{#1}{#2}}
\newcommand{\satisfaction}[2]{\outcome{S}{#1}{#2}}
\newcommand{\security}[2]{\outcome{Sec}{#1}{#2}}

\newcommand*\property[1]{\ifcase#1 \Circle\or\LEFTcircle\or\CIRCLE\fi}

\newacronym{AES}{AES}{Advanced Encryption Standard}
\newacronym{API}{API}{Application Programming Interface}
\newacronym{APS}{APS}{Analogical Problem Solving}
\newacronym{BEFKI GC-K}{BEFKI GC-K}{Short scale for \gls{Gc} of the Berliner Test to measure fluid and crystal intelligence}
\newacronym{BOMAT}{BOMAT}{Bochumer Matritzentest}
\newacronym{CBC}{CBC}{Cipher Block Chaining}
\newacronym{DES}{DES}{Data Encryption Standard}
\newacronym{ECB}{ECB}{Eletronic Codebook}
\newacronym{Gc}{\emph{Gc}}{Crystal Intelligence}
\newacronym{GCM}{GCM}{Galois/Counter Mode}
\newacronym{Gf}{\emph{Gf}}{Fluid Intelligence}
\newacronym{GQM}{GQM}{Goal Question Metric}
\newacronym{HEXACO}{HEXACO}{Honesty-Humility (H), Emotionality (E), Extraversion (X), Agreeableness (A), Conscientiousness (C), and Openness to Experience (O)}
\newacronym{IDE}{IDE}{Integrated Development Environment}
\newacronym{IQ}{IQ}{intelligence quotient}
\newacronym{IV}{IV}{initialization vector}
\newacronym{JDK}{JDK}{Java Development Kit}
\newacronym{nonce}{nonce}{number used once}
\newacronym{RPM}{RPM}{Raven's Progressive Matrices}
\newacronym{RS}{RS}{research subject}
\newacronym{USB}{USB}{Universal Serial Bus}
\newacronym{SHA}{SHA}{Secure Hash Algorithm}
\newacronym{SUS}{SUS}{System Usability Scale}
\newacronym{TLS}{TLS}{Transport Layer Security}
\newacronym{URL}{URL}{Uniform Resource Locator}

\setcopyright{rightsretained}

\copyrightyear{2020}
\acmYear{2020}
\setcopyright{rightsretained}
\acmConference[ICSE '20 Companion]{42nd International Conference on Software Engineering Companion}{October 5--11, 2020}{Seoul, Republic of Korea}
\acmBooktitle{42nd International Conference on Software Engineering Companion (ICSE '20 Companion), October 5--11, 2020, Seoul, Republic of Korea}\acmDOI{10.1145/3377812.3390892} 
\acmISBN{978-1-4503-7122-3/20/05}



\begin{document}
	\title[Effects of Code Examples on the Usability of Crypto APIs]{Fluid Intelligence Doesn't Matter! Effects of Code Examples on the Usability of Crypto APIs}

%
\author{Kai Mindermann}
\orcid{0000-0001-5155-9179}
\affiliation{%
	\institution{IC Consult}
	\city{Stuttgart}
	\country{Germany}
}
\email{kai.mindermann@ic-consult.com}

\author{Stefan Wagner}
\orcid{0000-0002-5256-8429}
\affiliation{%
	\institution{University of Stuttgart}
	\city{Stuttgart}
	\country{Germany}}
\email{stefan.wagner@iste.uni-stuttgart.de}
%
%
%
%
%
%


	\begin{abstract}
	\emph{Context}: 
	Programmers frequently look for the code of previously solved problems that they can adapt for their own problem.
	Despite existing example code on the web, on sites like Stack Overflow, cryptographic \glspl{API} are commonly misused. 
	There is little known about what makes examples helpful for developers in using crypto \glspl{API}.
	Analogical problem solving is a psychological theory that investigates how people use known solutions to solve new problems. 
	There is evidence that the capacity to reason and solve novel problems a.k.a \gls{Gf} and structurally and procedurally similar 
	solutions support problem solving.
	\emph{Aim}: 
	Our goal is to understand whether similarity and \gls{Gf} also have an effect in the
	context of using cryptographic \glspl{API} with the help of code examples.
	\emph{Method}: 
	We conducted a controlled experiment with 76 student participants developing with or without procedurally similar examples, one of two Java crypto libraries and measured the \gls{Gf} of the participants as well as the effect on usability (effectiveness, efficiency, satisfaction) and security bugs.
	\emph{Results}: 
	We observed a strong effect of code examples with a high procedural similarity on all dependent variables.
	Fluid intelligence \gls{Gf} had no effect. 
	It also made no difference which library the participants used.
	\emph{Conclusions}: 
	Example code must be more highly similar to a concrete solution, not very abstract and generic to have a positive effect in a development task. 
\end{abstract}

	\maketitle

	\glsresetall

\section{Introduction}
Most software requires the use of cryptography for security purposes.
Developers, who commonly are no security experts, must find out how to use cryptography \glspl{API} in their code.
Finding and learning from examples for the usage of such APIs is one of the ways to make their programs work. Yet, there is little known about what makes examples from the documentation or sites like Stack Overflow helpful for developers or lead to misuse.


Such an approach falls under  \emph{\gls{APS}} \cite{Gick1980}
in psychology.
There are three dimensions of similarity between a target solution and a source analogy \cite{Chen2002}: 
\begin{enumerate*}
	\item \emph{Superficial similarity} is given if the source analogy shares common general attributes like objects and characters. 
	\item \emph{Structural similarity} is given if the source analogy shares causal relations like the same solution principle, the same obstacles, the same outcome or the same resources.
	\item \emph{Procedural similarity} regards how similar the procedures and operational details are that are required to implement the solution.
\end{enumerate*}
Having only superficial similarity can lead to choosing unsuitable solutions. 
Hence, structural and procedural similarity are needed to support \gls{APS}.
Yet, not only the analogy itself is a factor but also the capabilities of the developer who relies on analogies to solve a problem. The most relevant capability is \gls{Gf} -- the ability to reason and solve novel problems -- because of its high correlation with analogical reasoning performance~\cite{Morrison2001}.

We designed a controlled experiment to test whether examples with different similarity and \emph{Gf} of developers have an
influence on the effectiveness, efficiency, satisfaction and created security bugs of performing a Java development task involving a crypto library done
by novice programmers. 
Examples from \emph{CryptoExamples}\footnote{\url{https://www.cryptoexamples.com/}} are our benchmark 
and were given the experiment group in addition to both groups being able to search the web for other examples.
To control for the influence of a specific crypto library, we included Google Tink besides the \gls{JDK}.

\citeauthor{Acar2017} \cite{Acar2017} conducted a controlled experiment with 256 Python developers who had to perform various tasks involving symmetric and asymmetric encryption. 
Their findings suggest that missing documentation, code examples and other functionality let participants struggle with the \glspl{API}.
\citeauthor{Mindermann2018} \cite{Mindermann2018} looked at the usability of Rust cryptography \glspl{API}. 
They found that the crypto library designed to be more usable was slightly less usable for the experiment participants. 
A major complaint by the participants was missing documentation of and examples for the libraries. 
In summary, low-level APIs, problems with documentation and missing examples are reasons for misuse.

\section{Experimental Design}

We had two \emph{independent variables} that we manipulated in the experiment: 
The used cryptographic library (either \gls{JDK} or Tink) and the used examples (either \emph{CryptoExamples} or other found examples). 
Additionally, we measured the \emph{independent variable} \gls{Gf} with \gls{BOMAT} advanced short~\cite{Hossiep2001}, a non-verbal power-speed test.

The \emph{dependent variables} are: 
The effectiveness (how much of the task was completed until the time limit was reached), efficiency (the time needed to to finish the task or, in case of not finishing the complete task, 80 minutes) and the satisfaction (measured with the \gls{SUS} \cite{Brooke1996}). Furthermore, we measured the number of statically detectable security bugs in the final implementation.


The subjects of the experiment were 76 undergraduate students of a course on the introduction to software engineering at
the University of Stuttgart, Germany. 
Participants were assigned to the following groups:
\begin{enumerate*}
	\item[(JO)] Using the default crypto library of the \gls{JDK} and not receiving the crypto example code.
	\item[(JM)] Using the default crypto library of the \gls{JDK} and receiving the crypto example code.
	\item[(TO)] Using the Tink crypto library and not receiving the crypto example code.
	\item[(TM)] Using the Tink crypto library and receiving the crypto example code.
\end{enumerate*} 

The only experiment task was to encrypt a string using the \gls{AES} and decrypt it for the tamper verification afterwards. 
Important to note here is, that none of the examples from the web (including \emph{CryptoExamples}) could be copied without substantial alteration to solve the task.

\section{Analysis}

\emph{Effectiveness.}
We calculated the following mean effectiveness for each group: \effectiveness{JO}{0.376}, \effectiveness{JM}{0.836}, \effectiveness{TO}{0.368} and \effectiveness{TM}{0.841}.
By comparing these values we could see a highly increased effectiveness for the participants who used the provided examples. 
Interestingly, Tink had a slightly worse effectiveness than the \gls{JDK} for the participants who did not get the code examples.



\emph{Efficiency.}
The groups had \efficiency{JO}{0.298}, \efficiency{JM}{0.993}, \efficiency{TO}{0.348} and \efficiency{TM}{1.287}.
There we could see a similar relation between the groups as previously for the effectiveness.
In terms of tasks per hour, participants were much more efficient on average (\SI{269}{\percent}) with the provided examples.
If we compare only group J and T, Tink users were also much more efficient on average (\SI{65}{\percent}).


\emph{Satisfaction.}
A general average score on the \gls{SUS} is said to be $68$ which none of the experiment groups reached: \satisfaction{JO}{39.7}, \satisfaction{JM}{58.1}, \satisfaction{TO}{46.6} and \satisfaction{TM}{61.6}. 
The values suggest that users of Tink might be a little more satisfied with the library than users of the \gls{JDK} and that the provided examples seem to increase the satisfaction as well and not only the effectiveness and efficiency.


\emph{Security Bugs.}
We found a mean number of security bugs of \security{JO}{3.28} and \security{JM}{0.33}.
%
Most bugs (40) were about \enquote{cipher with no integrity} followed by \enquote{\gls{ECB} mode is insecure} (30), \enquote{cipher is susceptible to Padding Oracle} (7) and \enquote{hard coded key} (4). 
All with \emph{high confidence} and a rank of \emph{scary}, the highest one in SpotBugs.

In a regression analysis of all involved factors, we found a negligible influence of \emph{Gf} with $\beta$ values close to 0. Only the usage of CrypotExamples showed statistically significant, large effects.

\section{Limitations}

Allowing the participants to use the web restricts our control on what sources they use to help them in their tasks. 
Participants of all groups could use the web and search for examples on the web, including the experiment group.
We see it as necessary, because it is closer to development in a practical setting. 
Yet, it introduces the threat that participants from the control group used examples with a procedural similarity comparable to \emph{CryptoExamples} (or \emph{CryptoExamples} directly). To mitigate this threat
\begin{enumerate*}
	\item we classified the other viewed examples and compared them to the provided examples from \emph{CryptoExamples}.
	\item After the experiment, we reassigned participants based on their activity log (if they found and used \emph{CryptoExamples}).
\end{enumerate*}
During this discovery, we found no other used example that has a comparable procedural similarity as \emph{CryptoExamples}.
Differences in other categories, except for security, are not as clear as for procedural similarity. 

We used first and second year students with some programming experience and little to no experience with cryptography. Hence, we expect that the results can be generalized to other programming beginners or even professionals new to using cryptography libraries. 

\section{Conclusions}




Our results suggests that providing analogical solution with high procedural similarity can be an important factor for being effective, efficient and satisfied in using cryptographic libraries while creating few security bugs.
%
%
To our surprise, the effect of \gls{Gf} was very small.
%
%
%
We see this positive, because it seems that successful usage of the cryptographic libraries with the help of examples does not depend on the \gls{Gf} of the developers. This means that there is no internal and stable factor that prevents developers from successful use. Improving solution examples is probably much easier than improving the fluid intelligence \gls{Gf} of developers.

\end{document}